\documentclass[aps,prd,twocolumn,superscriptaddress,amsfont,amssymb,amsmath,nofootinbib,showpacs,balancelastpage]{revtex4-1}

\usepackage{graphicx,longtable,natbib,mathrsfs,xcolor,ulem,array,float}
\usepackage{txfonts,subfigure,aas_macros, amsmath}
\usepackage{hyperref}

\newcommand{\bs}{\boldsymbol}

\newcommand{\lb}{\left\langle}
\newcommand{\rb}{\right\rangle}

\begin{document}

\addtolength{\hoffset}{-0.525cm}
\addtolength{\textwidth}{1.05cm}
\title{Probing Primordial Chirality in the Matter Distribution}

\author{Fang-Na Shao}
\affiliation{Department of Astronomy, Xiamen University, Xiamen, Fujian 361005, China}

\author{Hao-Ran Yu}\email{haoran@xmu.edu.cn}
\affiliation{Department of Astronomy, Xiamen University, Xiamen, Fujian 361005, China}

\author{Ming-Jie Sheng}
\affiliation{Department of Astronomy, Xiamen University, Xiamen, Fujian 361005, China}

\author{Bing-Hang Chen}
\affiliation{Department of Astronomy, Xiamen University, Xiamen, Fujian 361005, China}

\author{Huiyuan Wang}
\affiliation{Department of Astronomy, University of Science and Technology of China, Hefei, Anhui 230026, China}
\affiliation{School of Astronomy and Space Science, University of Science and Technology of China, Hefei, Anhui 230026, China}

%\date{\today}

\begin{abstract}
Whether parity symmetry was violated in the early universe remains one of the fundamental open questions in cosmology. If so, it may leave an intrinsic handedness in the large-scale matter distribution. Here we formulate a helicity-based estimator to measure this handedness. Using cosmological simulations with parity-violating initial conditions, we show that it survives nonlinear structure formation. Applying the estimator to density fields reconstructed from the SDSS DR7 galaxy catalog, we find mild deviations but statistically insignificant evidence for parity violation in the local universe. Our results establish the intrinsic handedness of the matter distribution as an observable relic of primordial parity violation, enabling a direct probe with current and future large-scale structure surveys.
\end{abstract}
\vspace{-16pt}

\maketitle

%--------------------------------- This part is introduction----------------------------------
\setlength{\parskip}{0pt}
The fundamental laws of nature were long presumed to be governed by elegant symmetries. This paradigm shifted dramatically with the discovery that parity and charge-conjugation symmetries were explicitly violated by the electroweak interaction \citep{lee1956QuestionParity}. This raises the question of whether microscopic parity violation could be transmitted to macroscopic scales during the inflationary epoch. Specifically, theories suggest that if parity-violating fields are coupled to the inflaton, such interactions can imprint chiral signatures onto primordial curvature perturbations that survive as observable fossils \citep{lue1999CosmologicalSignature, soda2011ParityViolation, jeong2012ClusteringFossils, cabass2023CollidingGhosts, fujita2024ParityviolatingScalar}. Testing fundamental parity symmetry in the large-scale structure (LSS) therefore provides a unique window into new inflationary physics.

Current searches for parity violation span various cosmological observables. 
Tensorial signatures are probed through their distinctive imprints on CMB polarization \citep{saito2007ProbingPolarization,contaldi2008AnomalousCosmicMicrowaveBackground} and the three-dimensional LSS \citep{jeong2012ClusteringFossils,masui2017TwoThreeDimensional, zhu2025SystematicAnalysis}, typically sourced by primordial gravitational waves from inflationary couplings \citep{jackiw2003ChernSimonsModification, satoh2008CircularPolarization}.  
Complementary searches for vectorial signatures focus on helically asymmetric fossils that manifest observationally through cosmic angular momenta induced by primordial tidal torques \citep{iye2019SpinParity,yu2020ProbingPrimordial,motloch2020ObservationalDetection, motloch2022ObservationalSearch,sheng2022SpinConservationcosmic,sheng2023BaryonicEffects,shim2025ProbingVector,sheng2025ObservationalEvidence}. Scalar signatures are explored through the parity-odd four-point correlation function (4PCF) and trispectrum \citep{philcox2022ProbingParity, hou2023MeasurementParityodd, coulton2024SignaturesParityviolatinga}, as predicted in inflationary scenarios involving Chern-Simons and axion-gauge interactions \citep{fujita2024ParityviolatingScalar,cabass2023ParityViolation,creque-sarbinowski2023ParityviolatingTrispectrum}.

Regardless of the specific mechanism driving parity violation during inflation, 
%\tcr{whether it manifests in tensor, vector, or scalar perturbations,} 
it may leave parity-violating imprints on the underlying matter distribution \citep{jeong2012ClusteringFossils, cabass2023CollidingGhosts, masui2017TwoThreeDimensional, fujita2024ParityviolatingScalar}. This raises the question of whether such imprints are encoded as a helicity asymmetry in the intrinsic morphology of the matter distribution itself. If so, parity violation could be characterized through an imbalance between left- and right-handed helicity modes, providing a physically transparent probe of cosmic chiral structures.

In this Letter, we investigate the survival of helicity asymmetry encoded in the morphology of the matter distribution under nonlinear gravitational evolution. To enable observational searches, we develop a framework for probing primordial parity violation through the handedness of reconstructed density fields, built around a helicity-based quadratic estimator. We apply the estimator to SDSS-reconstructed density fields, including both low-redshift density fields and reconstructed initial conditions.
%-------------------------------------- Method ---------------------------------------------

\textit{Parity-violating initial conditions ---} 
While the detailed parity-violating signatures depend on the underlying inflationary model, we capture their common phenomenology using a scale-dependent helicity asymmetry. We therefore model this asymmetry by introducing controlled helical perturbations into otherwise standard cosmological initial conditions at characteristic scales. 
The resulting parity-violating initial density field is given by
\begin{equation}
	\delta_{\mathrm{tot}}(\bs x) = \delta_{\rm sym}(\bs{x}) + \delta_{\mathrm{hel}}(\bs{x}),
\end{equation}
where $\delta_{\rm sym}(\bs{x})$ denotes the standard overdensity field and $\delta_{\mathrm{hel}}(\bs{x})$ encodes the helically asymmetric component. Here $\delta_{\rm sym}(\bs{x})$ is generated from a Gaussian random field with linear power spectrum $P_{\rm lin}(k)$,
\begin{equation}
	\langle \tilde{\delta}_{\rm sym}(\bs{k}) \tilde{\delta}_{\rm sym}^*(\bs{k}') \rangle = (2\pi)^3 \delta_{\rm D}^{(3)}(\bs{k} - \bs{k}') P_{\rm lin}(k),
\end{equation}
constructed following the standard approach from Gaussian white noise $\tilde{\mu}(\bs{k})$ scaled by $\sqrt{P_{\rm lin}(k)}$. By construction, this overdensity field preserves statistical parity symmetry and contains no physical parity-violating signal.

To introduce a controlled parity-violating component, we exploit a geometric property of proto-filamentary structures. As one of the most prominent geometric features of the cosmic web, filaments are intrinsically unoriented curves, yet their torsion remains a pseudoscalar that changes sign under parity transformations. This makes torsion a natural carrier of chirality in the large-scale matter distribution. We therefore construct a helically asymmetric component $\delta_{\mathrm{hel}}(\bs{x})$ from an ensemble of helical proto-filamentary structures sharing a common handedness.
Specifically, we generate an ensemble of helices with radius $R$ and pitch $p$, parameterized as $\bs{r}(t)=(R\cos t,\,R\sin t,\,pt)$. These helices are randomly distributed in both position and orientation to satisfy the cosmological principle, but share the same sign of the pitch parameter $p$ throughout the simulation volume. The resulting field is subsequently smoothed with a Gaussian kernel. The coherent chirality of the helical ensemble globally violates parity symmetry.

We then rescale the power spectrum of the total overdensity field in Fourier space,
\begin{equation}
\tilde{\delta}_{\rm asym}(\bs{k}) = \left[ \dfrac{P_{\rm lin}(k)}{\langle |\tilde{\delta}_{\mathrm{tot}}(\bs{k})|^2 \rangle} \right]^{1/2} \tilde{\delta}_{\mathrm{tot}}(\bs{k}),
\end{equation}
ensuring that the resulting initial density field $\tilde{\delta}_{\rm asym}(\bs{k})$ is consistent with the linear power spectrum $P_{\rm lin}(k)$ given by the standard cosmological model. The resulting initial density field remains approximately Gaussian after this construction. Since the parity-violating component is encoded solely in the helical overdensity $\delta_{\mathrm{hel}}(\bs{x})$, parity asymmetry can be systematically introduced on any scale by varying the seeding scale of the helices.

We numerically generate these density perturbations and evolve them to late times using $N$-body simulations. A periodic box of $L=1280\,{\rm Mpc}\,h^{-1}$ per side is chosen to match a typical next-generation galaxy survey. 
For the parity-violating configurations, the helical overdensity field $\delta_{\mathrm{hel}}(\bs{x})$ is constructed from an ensemble of 200 helices with random positions and orientations, smoothed on a scale of $10\,{\rm Mpc}\,h^{-1}$. The helix lengths are uniformly distributed between $0$ and $L$, while the pitch $p$ is sampled from a Gaussian distribution with mean $L/6$ and standard deviation $L/200$. These helices thus share the same chirality and collectively contribute to a parity-violating signal predominantly seeded around the characteristic wave number $6k_{\rm f}$, where $k_{\rm f}=2\pi/L$ is the fundamental mode of the simulation box.

Parity-violating initial conditions are derived from the density field $\delta_{\mathrm{asym}}(\bs{x})$, with particle positions and velocities initialized via the Zel'dovich approximation at redshift $z=100$ using $N_p=512^3$ dark matter particles on a $512^3$ grid. These particles are subsequently evolved to redshift $z=0$ using the $N$-body code CUBE2 \citep{yu2026CUBE2Parallel}. To control cosmic variance and the stochasticity of the asymmetric components, we perform an ensemble of 500 independent simulations with helically asymmetric initial conditions. For rigorous comparison, we additionally evolve a matched control ensemble of 500 simulations without parity violation.

\textit{Helical asymmetry estimators ---} Anisotropic clustering in the density field encodes long-wavelength perturbations through scale-dependent mode coupling \citep{2012arXiv1202.5804P}. This coupling is characterized by the tidal shear tensor, which decomposes under rotations about the large-scale mode into scalar, vector, and tensor helicity modes. The scalar and tensor components have been extensively used in tidal reconstruction \citep{2012arXiv1202.5804P,zhu2016CosmicTidal,karacayli2019AnatomyCosmic}. Of particular relevance here, the vector sector admits an imbalance between left- and right-handed helicity modes that transform as pseudoscalars under parity, enabling a sensitive probe of parity-violating anisotropic mode coupling \citep{zhu2025SystematicAnalysis,bartolo2015ParityviolatingAnisotropic,jamieson2024ParityoddPowera}. Such signals are naturally captured by quadratic estimators.

Concretely, building on the quadratic mode-coupling framework introduced by Jeong and Kamionkowski \citep{jeong2012ClusteringFossils}, we construct a helicity-sensitive estimator to directly probe the helicity asymmetry of the density field. The estimator isolates the parity-odd helicity combination from mode-mode correlations, making it directly sensitive to handedness encoded in the density field, independent of the specific parity-violating mechanism. The helicity asymmetry of $\delta_{\mathrm{hel}}$ sources a nonvanishing parity-violating signal, which we characterize through the $N$-body ensemble described below. We therefore probe this asymmetry through the helicity-projected quadratic estimator,
\begin{equation}
	v_{\alpha}^{L/R}(\boldsymbol{K})=\mathcal{P}^{L/R}_{\alpha \beta}(\boldsymbol{K}) \sum_{\bs{k}} -i K_{\gamma} k_{\beta} (K-k)_{\gamma}  \tilde{\delta}(\bs{k}) \tilde{\delta}(\bs{K}-\bs{k}),
	\label{eq:estimator}
\end{equation}
where $\mathcal{P}^{L/R}$ denotes the two projection operators onto the  left- and right-handed helicity bases. 
This projection relies on the basic decomposition of a three-dimensional vector field: its three independent degrees of freedom are one longitudinal irrotational $E$-mode component, and two transverse divergence-free $L$- and $R$-mode components \citep{yu2020ProbingPrimordial}. The latter encode helicity information and are naturally represented in the left- and right-handed helicity bases, whose basis vectors are eigenvectors of the curl operator. The corresponding projection operators are
\begin{align}
    {\mathcal{P}}^E_{\alpha \beta} &\equiv  k_\alpha k_\beta/k^2,\\
    {\mathcal{P}}^{L/R}_{\alpha \beta} &\equiv  \dfrac{1}{2}\left( \delta_{\alpha \beta}-k_{\alpha}k_{\beta}/k^2 \pm i \epsilon_{\alpha \beta \gamma}k_{\gamma}/k   \right).
\end{align}
We define the corresponding helicity power spectra as
\begin{equation}
	\lb {\bs v}^{L/R}(\bs{k}) \cdot {\bs v}^{L/R \ast}(\bs{k}') \rb  \equiv  (2\pi)^3\delta_{\rm D}^{(3)}(\bs k-\bs k')P_{L/R}(k).
\end{equation}
To quantify the strength of the parity asymmetry, the helicity asymmetry parameter $\chi(k)$ is defined as
\begin{equation}
	\chi(k) \equiv \dfrac{P_L(k)-P_R(k)}{P_L(k)+P_R(k)}.
\end{equation}
Since the estimator ${\bs v}^{L/R}$ is constructed as a quadratic function of $\delta$, the helicity asymmetry $\chi(k)$ probes the parity-odd sector of the connected 4PCF of the density field. Instead of classifying geometric configurations according to their parity properties, $\chi$ directly measures the imbalance between left- and right-handed helicity modes. Under perfect parity symmetry, the left- and right-handed power spectra satisfy $P_L=P_R$, implying $\chi=0$. 
In the maximally parity-violating case, $\chi = \pm 1$, the perturbations are fully polarized into purely left- or right-handed helicity states. For the proto-filamentary initial conditions considered here, the measured helicity asymmetry statistically reflects the torsional handedness encoded in the injected helices. In practice, finite-volume realization noise induces statistical fluctuations of $\chi$ around zero even in the symmetric simulations, resulting in a nonzero variance. To quantify the statistical significance of the measured asymmetry, we combine the uncertainties from the symmetric and asymmetric realizations, denoted by $\sigma_{\mathrm{sym}}$ and $\sigma_{\mathrm{asym}}$, and define a joint uncertainty as $\sigma_{\mathrm{joint}} = \sqrt{\sigma_{\mathrm{sym}}^2 + \sigma_{\mathrm{asym}}^2}$. The detection significance is then quantified relative to $\sigma_{\mathrm{joint}}$.

\begin{figure}[t]
	\centering	\includegraphics[width=1\linewidth]{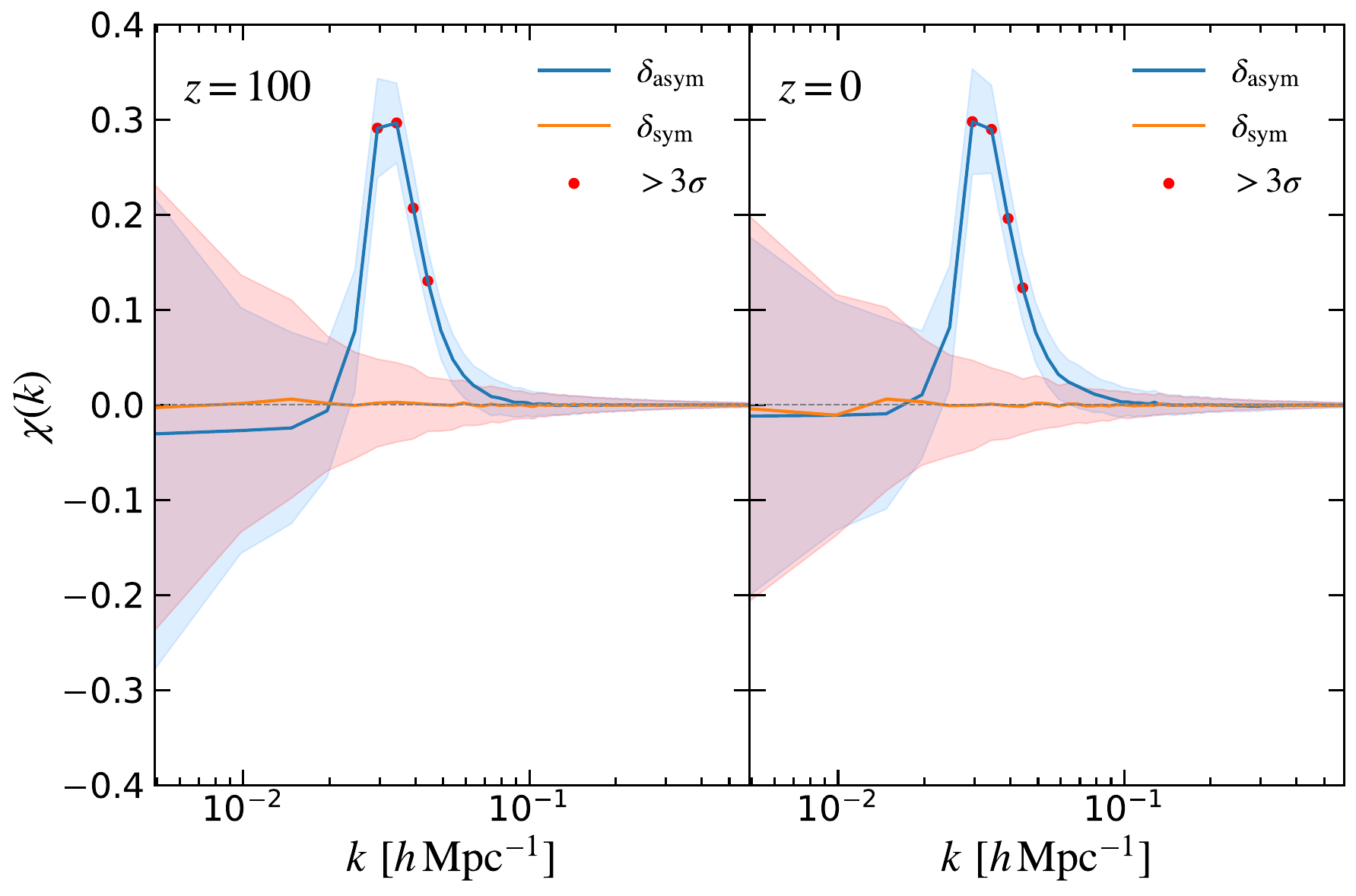}
	\caption{Helicity asymmetries of the primordial (left panel) and late-time (right panel) density fields for parity asymmetric (blue curves) and symmetric (red curves) simulations. 
        %The blue curves correspond to the asymmetric case which encodes the chiral information, while the orange curves represent the symmetric case. 
        The shaded regions indicate the standard deviation ($1\sigma$) of the ensemble-average. The red markers highlight the asymmetries exceeding a statistical significance of $3\sigma$.}\label{fig1}
\end{figure}

Fig.~\ref{fig1} shows the helicity asymmetries measured from an ensemble of 500 paired asymmetric and symmetric simulations evolved from $z=100 $ to $z=0$. At high redshift, asymmetric realizations exhibit a pronounced helicity signal peaking at the injection scale $k \simeq 6k_{\rm f}\simeq 0.029\,h\,{\rm Mpc}^{-1}$, with a statistical significance of $7.1\sigma$. %This signal remains stable throughout nonlinear evolution. 
At $z=0$, the significance remains at $6.7\sigma$, and the peak amplitude is nearly unchanged with $\chi_{\rm peak} \simeq 0.29$ at both epochs. In contrast, the symmetric control realizations are consistent with $\chi \simeq 0 $ across all scales, as expected since gravitational evolution induces no spurious helicity. 
As further consistency checks, reversing the injected handedness flips the sign of $\chi$, and varying the helix pitch shifts the peak to the corresponding scale. %(see Supplemental Material). 
This persistence from early to late times demonstrates that the helicity asymmetry captured by Eq.\eqref{eq:estimator} is robust against nonlinear gravitational evolution.

\textit{Observational examination ---} We further apply the analysis to an observationally constructed density field derived from the SDSS DR7 galaxy distribution within the ELUCID framework \citep{wang2016ELUCIDEXPLORING}. The field is constructed from galaxies in the Northern Galactic Cap over the redshift range $0.01 \le z \le 0.12$, where the underlying matter field is inferred from galaxy groups using the halo-domain method with redshift-space distortion corrections \citep{wang2009ReconstructingCosmic,wang2013RECONSTRUCTINGINITIAL}. The constructed density field spans a cubic volume of side length $L=675\,{\rm Mpc}/h$, is smoothed on $1\,{\rm Mpc}/h$, and incorporates the survey geometry through a sky-weight function with boundary regions excluded.

\begin{figure}[t]
	\centering\includegraphics[width=1\linewidth]{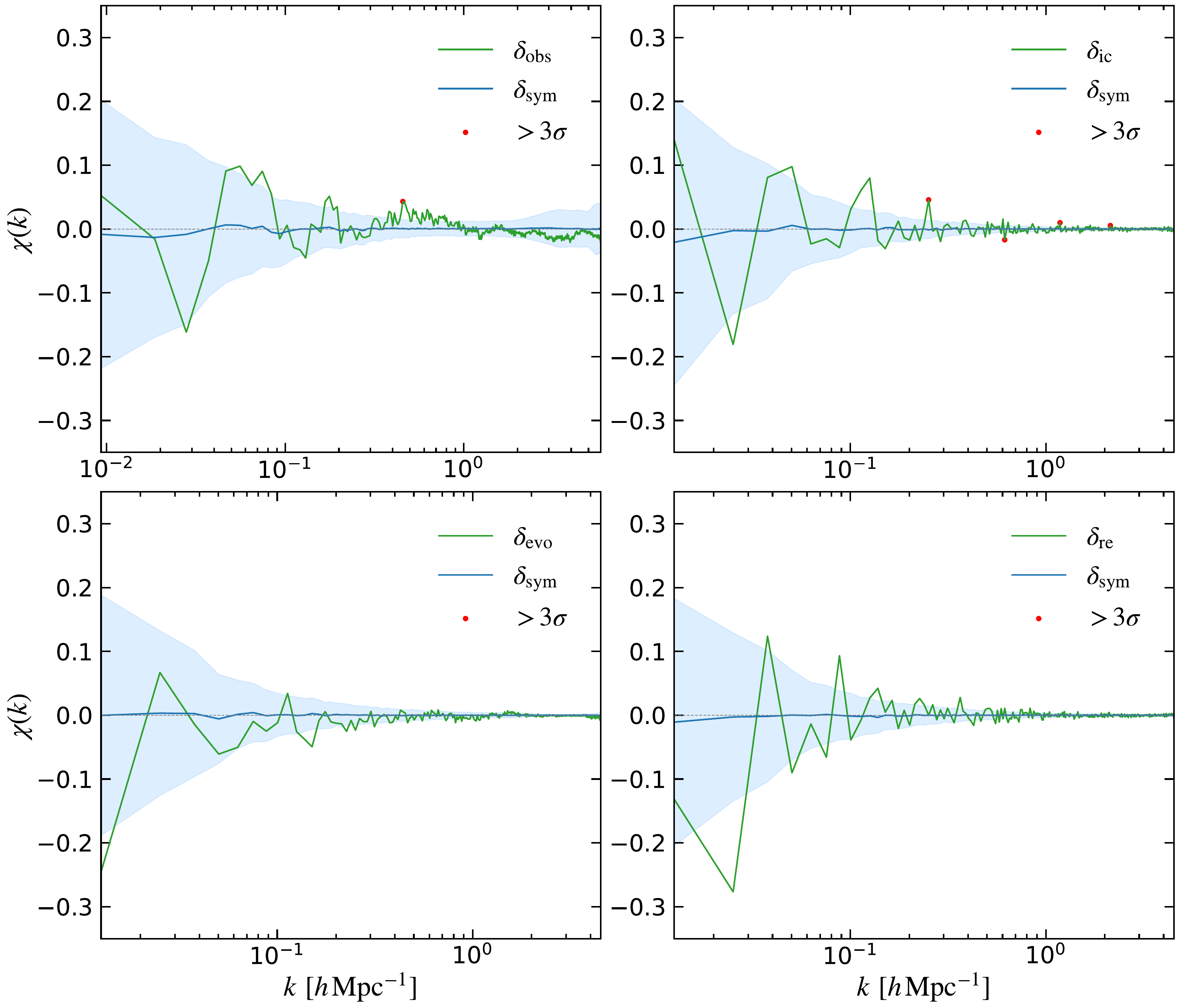} 
	\caption{Helicity asymmetry measurements of the local universe. The upper-left panel show the asymmetry of the estimated density field of local universe. The upper-right panel corresponds to the reconstructed initial condition. The lower panels show the asymmetry measurements of the density field evolved from the reconstructed initial condition, and the displacement field associated with this reconstructed cosmic evolution. The green curves denote the respective asymmetry measurements. The blue shaded regions indicate the $1\sigma$ confidence level from the corresponding symmetric realizations.}\label{fig2}
\end{figure}

We apply the method to the constructed SDSS density field to search for parity-violating signatures in the observed matter distribution. The top-left panel of Fig.~\ref{fig2} presents the helicity asymmetry measured from the ELUCID-constructed density field $\delta_{\rm obs}$ derived from the SDSS catalog. To assess the statistical significance, we generate 500 symmetric $N$-body simulations with $N_p=768^3$ dark matter particles in a box of size $L=675\,{\rm Mpc}/h$, applying the same survey mask and smoothing scale as used in the observational construction. Compared with the symmetric realizations, the observed signal is broadly consistent with the null hypothesis, remaining within the $1\sigma$ scatter over most scales. A localized marginal excess is present around $k \simeq 0.45\,h\,{\rm Mpc}^{-1}$, with a maximum significance of $3.04\sigma$. We therefore find mild deviations but no statistically significant evidence for parity-violating imprints in the SDSS-reconstructed low-redshift density field.

We extend the analysis to the reconstructed initial conditions of the local universe within the ELUCID framework \citep{wang2016ELUCIDEXPLORING,wang2014ELUCIDEXPLORING}. The reconstruction employs a Hamiltonian Markov Chain Monte Carlo algorithm combined with Particle-Mesh dynamics to infer the initial density field consistent with the SDSS DR7 observations. The remaining panels of Fig.~\ref{fig2} present the helicity asymmetry measured from three fields derived from this reconstruction --- 
(i) the reconstructed initial density field $\delta_{\rm ic}$, 
(ii) the evolved late-time density field $\delta_{\rm evo}$ obtained by evolving $\delta_{\rm ic}$ to $z=0$, 
and (iii) the recovered linear density field $\delta_{\rm re}$ reconstructed from the Lagrangian displacement field through $\delta_{\rm re} \propto -\nabla \cdot {\bs \Psi}(\mathbf{q})$. The recovered density field $\delta_{\rm re}$ traces the compressional $E$-mode component of the Lagrangian displacement field and preserves more primordial information than the nonlinear Eulerian density field \citep{yu2017NonlinearMode}. All fields are defined in a volume of side length $L = 500 \,\mathrm{Mpc}/h$ with grid resolution $500^3$. To quantify the statistical significance, we compare against a control ensemble of 500 independent symmetric realizations generated with the same volume and resolution.

We find that the helicity signals measured from these reconstructed fields remain within the $1\sigma$ scatter of the symmetric realizations. Deviations are localized and statistically limited; in particular, the reconstructed initial density field exhibits a maximum significance of $3.32\sigma$ around $k \simeq 1.18 \, h\,{\rm Mpc}^{-1}$. However, given the absence of coherent broad-band features and the consistency of both the evolved and recovered fields with the null hypothesis, no significant evidence is found for parity-violating imprints in the SDSS-reconstructed initial conditions.

%--------------------------------------Conlusion and Discussion----------------------------------------

\textit{Conclusion and discussion\,---\,}
We have shown that helicity asymmetry encoded in the intrinsic handedness of the large-scale matter distribution survives nonlinear gravitational evolution, remaining detectable even in the strongly nonlinear regime at $z=0$. Using a helicity-based quadratic estimator together with phenomenological parity-violating initial conditions, we establish a framework for probing primordial parity violation that is independent of any specific inflationary mechanism and broadly applicable to scenarios predicting parity-violating or scale-dependent chiral imprints. These results demonstrate that helicity asymmetry provides a robust signature of primordial parity violation preserved throughout nonlinear structure formation.

As a first observational application, we applied this framework to ELUCID-reconstructed density fields derived from the SDSS DR7 galaxy catalog. Some reconstructed fields exhibit localized excesses reaching $\sim3\sigma$. Unlike the injected signals in our simulations, these excesses occur at different scales and are not consistently reproduced across reconstructed fields. The absence of coherent broadband features prevents these excesses from being identified as robust parity-violating signatures, and we find mild deviations but statistically insignificant evidence for primordial parity violation in the local universe. We note, however, that the strong Gaussian and power-spectrum priors imposed during reconstruction \citep{wang2014ELUCIDEXPLORING} may systematically suppress genuine primordial signals, highlighting the importance of reconstruction methods that better retain parity-sensitive information.
 
The intrinsic handedness of the matter distribution provides a new observable of primordial parity violation, complementing existing searches based on CMB polarization \citep{saito2007ProbingPolarization,contaldi2008AnomalousCosmicMicrowaveBackground}, galaxy angular momenta \citep{iye2019SpinParity,yu2020ProbingPrimordial,motloch2020ObservationalDetection, motloch2022ObservationalSearch,shim2025ProbingVector}, and parity-odd higher-order statistics \citep{philcox2022ProbingParity, hou2023MeasurementParityodd, coulton2024SignaturesParityviolatinga}. Since the estimator operates directly on reconstructed matter fields without relying on a specific tracer, the approach is broadly applicable to galaxy surveys, H\,{\sc i} intensity mapping, and weak lensing, enabling future cross-tracer consistency tests of candidate signals. Beyond establishing the existence of parity asymmetry, this framework opens a route to characterizing its scale dependence and exploring its connection with the morphology of the cosmic web and parity-violating mechanisms in the early universe. Realizing these prospects will require three-dimensional reconstructions that preserve parity-sensitive information with reduced prior dependence. Upcoming surveys will provide the volume needed to fully exploit such reconstructions and substantially improve sensitivity to parity-violating imprints in the large-scale structure of the Universe.

\textit{Acknowledgements ---} We thank Ue-Li Pen for insightful discussions and valuable suggestions throughout this work. We acknowledge the support from the National Natural Science Foundation of China (NSFC) No. 124B2054, 12595312, 12192224, and the CAS Project for Young Scientists in Basic Research, No. YSBR-062.

\bibliographystyle{apsrev4-1}
\bibliography{fangna_ref}

\end{document}